\documentclass[11pt,fleqn]{article}
\usepackage{amsfonts, epsfig, amssymb, amsmath}
\usepackage[normalem]{ulem}
\usepackage{color}
\newcommand{\ol}{\setlength{\itemsep}{0pt.}\begin{enumerate}}
\newcommand{\eol}{\end{enumerate}\setlength{\itemsep}{-\parsep}}
\newcommand{\ignore}[1]{}
\title{On the OpenAI whole-cube bound}
\author{Alex Samorodnitsky}

\begin{document}
\date{}
\maketitle


\newtheorem{THEOREM}{Theorem}[section]
\newenvironment{theorem}{\begin{THEOREM} \hspace{-.85em} {\bf :}
}%
                        {\end{THEOREM}}
\newtheorem{LEMMA}[THEOREM]{Lemma}
\newenvironment{lemma}{\begin{LEMMA} \hspace{-.85em} {\bf :} }%
                      {\end{LEMMA}}
\newtheorem{COROLLARY}[THEOREM]{Corollary}
\newenvironment{corollary}{\begin{COROLLARY} \hspace{-.85em} {\bf
:} }%
                          {\end{COROLLARY}}
\newtheorem{PROPOSITION}[THEOREM]{Proposition}
\newenvironment{proposition}{\begin{PROPOSITION} \hspace{-.85em}
{\bf :} }%
                            {\end{PROPOSITION}}
\newtheorem{DEFINITION}[THEOREM]{Definition}
\newenvironment{definition}{\begin{DEFINITION} \hspace{-.85em} {\bf
:} \rm}%
                            {\end{DEFINITION}}
\newtheorem{EXAMPLE}[THEOREM]{Example}
\newenvironment{example}{\begin{EXAMPLE} \hspace{-.85em} {\bf :}
\rm}%
                            {\end{EXAMPLE}}
\newtheorem{CONJECTURE}[THEOREM]{Conjecture}
\newenvironment{conjecture}{\begin{CONJECTURE} \hspace{-.85em}
{\bf :} \rm}%
                            {\end{CONJECTURE}}
\newtheorem{MAINCONJECTURE}[THEOREM]{Main Conjecture}
\newenvironment{mainconjecture}{\begin{MAINCONJECTURE} \hspace{-.85em}
{\bf :} \rm}%
                            {\end{MAINCONJECTURE}}
\newtheorem{PROBLEM}[THEOREM]{Problem}
\newenvironment{problem}{\begin{PROBLEM} \hspace{-.85em} {\bf :}
\rm}%
                            {\end{PROBLEM}}
\newtheorem{QUESTION}[THEOREM]{Question}
\newenvironment{question}{\begin{QUESTION} \hspace{-.85em} {\bf :}
\rm}%
                            {\end{QUESTION}}
\newtheorem{REMARK}[THEOREM]{Remark}
\newenvironment{remark}{\begin{REMARK} \hspace{-.85em} {\bf :}
\rm}%
                            {\end{REMARK}}

\newcommand{\thm}{\begin{theorem}}
\newcommand{\lem}{\begin{lemma}}
\newcommand{\pro}{\begin{proposition}}
\newcommand{\dfn}{\begin{definition}}
\newcommand{\rem}{\begin{remark}}
\newcommand{\xam}{\begin{example}}
\newcommand{\cnj}{\begin{conjecture}}
\newcommand{\mcnj}{\begin{mainconjecture}}
\newcommand{\prb}{\begin{problem}}
\newcommand{\que}{\begin{question}}
\newcommand{\cor}{\begin{corollary}}
\newcommand{\prf}{\noindent{\bf Proof:} }
\newcommand{\ethm}{\end{theorem}}
\newcommand{\elem}{\end{lemma}}
\newcommand{\epro}{\end{proposition}}
\newcommand{\edfn}{\bbox\end{definition}}
\newcommand{\erem}{\bbox\end{remark}}
\newcommand{\exam}{\bbox\end{example}}
\newcommand{\ecnj}{\bbox\end{conjecture}}
\newcommand{\emcnj}{\bbox\end{mainconjecture}}
\newcommand{\eprb}{\bbox\end{problem}}
\newcommand{\eque}{\bbox\end{question}}
\newcommand{\ecor}{\end{corollary}}
\newcommand{\eprf}{\bbox}
\newcommand{\beqn}{\begin{equation}}
\newcommand{\eeqn}{\end{equation}}
\newcommand{\wbox}{\mbox{$\sqcap$\llap{$\sqcup$}}}
\newcommand{\bbox}{\vrule height7pt width4pt depth1pt}
\newcommand{\qed}{\bbox}

\def\H{\{0,1\}^n}

\def\S{S(n,w)}

\def\g{g_{\ast}}
\def\xop{x_{\ast}}
\def\y{y_{\ast}}
\def\z{z_{\ast}}

\def\f{\tilde f}

\def\n{\lfloor \frac n2 \rfloor}

\def \E{\mathop{{}\mathbb E}}
\def \R{\mathbb R}
\def \Z{\mathbb Z}
\def \F{{\cal F}}
\def \S{\mathbb S}

\def \P{{\cal P}}

\def \x{\textcolor{red}{x}}
\def \r{\textcolor{red}{r}}
\def \Rc{\textcolor{red}{R}}

\def \noi{{\noindent}}

\def \iff{~~~~\Longleftrightarrow~~~~}

\def\myblt{\noi --\, }

\def \queq {\quad = \quad}

\def\<{\left<}
\def\>{\right>}
\def \({\left(}
\def \){\right)}

\def \e{\epsilon}
\def \l{\lambda}

\def\myblt{\noi --\, }

\def\Tp{Tchebyshef polynomial}
\def\Tps{TchebysDeto be the maximafine $A(n,d)$ l size of a code with distance $d$hef polynomials}
\newcommand{\rarrow}{\rightarrow}

\newcommand{\larrow}{\leftarrow}

\overfullrule=0pt
\def\setof#1{\lbrace #1 \rbrace}

\begin{abstract}
This is an essentially derivative note, whose goal is to interpret the 'whole-cube' bound of OpenAI for binary codes in a possibly somewhat more accessible way. This is attained, in part, by connecting it to some previously known results. All of the heavy technical lifting in the interpretation below is also due to the OpenAI language models ChatGPT 5.6 and ChatGPT 6. With that, we hope that the statements of the main results, the ensuing discussion, and the arguments themselves may be of some interest. 

\end{abstract}

\section{Introduction}

\noi The goal of this note is to present an interpretation of one of the new OpenAI upper bounds (\cite{OpenAI}) on the asymptotic rate of binary codes with given minimum distance. We believe that the argument presented here might be somewhat more accessible. With that, it should be noted that, as opposed to the proof in \cite{OpenAI}, it is not self-contained, since it makes use of the results in \cite{AS}. {\it Our contribution}: As mentioned in the abstract, essentially all of the arguments below are due to the OpenAI language models ChatGPT 5.6 and ChatGPT 6, including a (version of) the statement of Theorem~\ref{thm:main}. Our main contribution is in observing that 'harmonic spaces' used in the argument of \cite{OpenAI} seem to be similar to eigenspaces of the adjacency matrix of the Hamming ball, as studied in \cite{AS}, and in badgering the language models to explain this connection. We also rewrote the proof of Theorem~\ref{thm:main} somewhat, to make it easier to understand (at least to this reader). 

\noi  We refer to the cited literature for background on coding theory and the rate–distance problem. Here let us only recall that a binary error-correcting code $C$ of length $n$ and minimum distance $d$ is a subset of the Hamming cube $\H$ in which the distance between any two distinct points is at least $d$. Let $A(n, d)$ be the maximal size of such a code.
The quantity
\[
R(\delta) ~=~ \limsup_{n \rarrow \infty} \frac 1n \log_2 A\(n,\lfloor \delta n \rfloor\),
\]
is the {\it asymptotic maximal rate} of the code with relative distance $\delta$, for $0 \le \delta \le \frac12$.

\thm
\label{thm:main}

Let $C$ be a binary code of length $n$ and minimum distance $d$. Let $B$ be a connected subset of $\H$ whose adjacency matrix has $t$ distinct positive eigenvalues $\l_1 > ...> \l_t$ with corresponding multiplicities $m_1,...,m_t$. Then for any $1 \le k \le t$ such that $\frac{\l^2_k}{\l_1} \ge n - 2d + 1$ we have
\[
|C| ~\le~ 2d \cdot \frac{|B|}{m_k}.
\]
\ethm

\noi We note that for the special case $k = 1$ this claim was stated and proved in \cite{FT}, in the case where $C$ is a linear code. It was extended to general binary codes in \cite{NS}. It was used in these papers to rederive the first linear programming bound of \cite{MRRW} for linear (correspondingly general) codes. 

\noi As a corollary, we rederive the OpenAI whole-cube bound for binary codes. For $0 \le \beta < \alpha \le \frac12$, let $\Gamma_H(\alpha,\beta) = \frac{2(\alpha-\beta)(1-\alpha-\beta)}{\sqrt{\alpha(1-\alpha)}}$. Let 
\[
\kappa_H(\delta) ~=~ \inf_{\begin{array}{ccc} 0 \le \beta < \alpha \le \frac12 \\ \Gamma_H(\alpha,\beta) > 1 - 2\delta\end{array}} \Big\{H(\alpha) - H(\beta)\Big\}, 
\]
where $H(x) = x \log_2\(\frac 1x\) + (1-x) \log_2\(\frac{1}{1-x}\)$ is the binary entropy function. 
Then
\cor 
\label{cor:WCB}
\[
R(\delta) ~\le~ \kappa_H(\delta).
\]
\ecor 

\noi {\bf Discussion: Faber-Krahn-type questions}.

\noi For $k = 1$, Theorem~\ref{thm:main} was proved in \cite{FT} (for binary linear codes). It was then observed that choosing the set $B$ to be a Hamming ball allows to recover the first bound of \cite{MRRW} on $R(\delta)$. The following question was then posed in \cite{FT}: For a given $0 < \alpha < 1$, what is the largest possible eigenvalue of a subset $B$ of $\H$ of cardinality $2^{\alpha n}$. This question was answered in \cite{LSob}, where the Hamming balls were shown to have essentially the largest eigenvalue for their size. Hence, in this special case,  a different choice of $B$ will not lead to a better bound on the asymptotic rate function. 

\noi Theorem~\ref{thm:main} as stated here leads naturally to more general questions about the spectra of induced subgraphs of the cube, in particular about the possible sizes of their eigenvalues and the corresponding multiplicities. It will be shown in the proof of Corollary~\ref{cor:WCB} that one can recover the whole-cube bound of \cite{OpenAI} from the theorem by choosing $B$ to be a Hamming ball. It is conceivable that one can improve the bound by a different choice of $B$.

\section{Proofs}

\subsection{Proof of Corollary~\ref{cor:WCB}}

\noi We apply Theorem~\ref{thm:main} to a Hamming ball, and use the results from \cite{AS} on its spectrum.

\noi In order to describe the results in \cite{AS}, let us introduce some notation. For a positive integer $m$, let the Krawtchouk polynomials $K^{(m)}_0,...,K^{(m)}_m$ be the family of polynomials orthogonal w.r.t. the binomial measure $\mu(i) = \frac{{m \choose i}}{2^m}$. For two integers $0 \le b \le a$ let $x_b$ denote the first root of the Krawtchouk polynomial $K^{(n-2b)}_{a-b+1}$. It was shown in \cite{AS} that for any $0 \le b \le a$ the adjacency matrix of the Hamming ball of radius $a$ has an eigenvalue $\l = (n-2b) - 2x_b$ whose multiplicity is at least $m = {n \choose b} - {n \choose {b-1}}$. 

\noi The asymptotics of the first roots of the Krawtchouk polynomials are known (see e.g., \cite{MRRW}). They imply that for any two constants $0 \le \beta < \alpha \le \frac12$, setting $a = \alpha n$ and $b = \beta n$ (and assuming w.l.o.g. that $a$ and $b$ are integers), we get that $\l = 2\sqrt{(\alpha - \beta)(1 - \alpha - \beta)} \cdot n + o(n)$. It is also shown in \cite{AS} that the maximal eigenvalue of the Hamming ball of radius $a$ is obtained by choosing $\beta = 0$ in this formula, that is $\l_1 = 2\sqrt{\alpha(1 - \alpha)} \cdot n + o(n)$ (this was already shown in \cite{FT}). It follows that $\frac{\l^2}{\l_1} = \frac{2(\alpha-\beta)(1-\alpha-\beta)}{\sqrt{\alpha(1-\alpha)}} \cdot n + o(n) = \Gamma_H(\alpha, \beta) \cdot n + o(n)$. Hence, for $d = \delta n$ and for a sufficiently large $n$, the condition $\Gamma_H(\alpha, \beta) > 1 - 2 \delta$ implies the condition $\frac{\l^2}{\l_1} \ge n - 2d + 1$ in Theorem~\ref{thm:main}, and we may apply the claim of the theorem, obtaining the upper bound $|C| \le n \cdot \frac{|B|}{m}$ for a code $C$ of distance $d$, where $B$ is the Hamming ball of radius $a = \alpha n$ and $m = {n \choose b} - {n \choose {b-1}}$, for $b = \beta n$. The claim of the corollary follows, varying $\alpha$ and $\beta$, and recalling the asymptotics of the binomial coefficients (\cite{Lint}: ${n \choose k} = 2^{n \cdot \(H\(k/n\) + o(1)\)}$. 

\eprf 

\subsection{Proof of Theorem~\ref{thm:main}}

\noi We begin with defining some relevant notions and describing their properties. 

\noi Let $A_B$ be the adjacency matrix of $B$. Let $\l_1$ be the maximal eigenvalue of $A_B$ and let $\phi$ be the corresponding strictly positive eigenfunction. Let $\l$ be an eigenvalue of $A_B$ such that $\frac{\l^2}{\l_1} \ge n-2d+1$. Let $m$ be the multiplicity of $\l$. Let $V$ be the $m$-dimensional eigenspace of $A_B$ corresponding to the eigenvalue $\l$. Define an inner product $\<\cdot,\cdot\>_{\phi}$ on $V$ as follows: For $f, g \in V$ let $\<f,g\>_{\phi} = \sum_{S \in B} \frac{f(S) g(S)}{\phi(S)}$. Let $\psi_1,...,\psi_m$ be an orthonormal basis of $V$ w.r.t. this inner product. Define a function $\F$ from $B$ to the cone ${\S}^m_+$ of positive semidefinite $m \times m$ matrices by setting for $S \in B$ 
\[
v(S) ~=~ \left(\begin{array}{c} \psi_1(S) \\ \psi_2(S) \\ \vdots \\ \psi_m(S) \end{array} \right) \quad \mathrm{and} \quad \F(S) = \frac{v(S) \otimes v(S)}{\phi(S)},
\]
where for $v,w \in \R^m$ we write $v \otimes w$ for the $m \times m$ matrix $v w^t = \(v_i w_j\)_{1 \le i,j \le m}$. 

\noi We extend $\F$ to a function from $\H$ to ${\S}^m_+$ by taking $\F(S) = 0$ for $S \not \in B$ (here we write $0$ for the zero matrix in ${\S}^m_+$). Let $A$ be the adjacency matrix of the cube $\H$. We define a function $A \F$ from $\H$ to ${\S}^m_+$ naturally by $(A \F)(x) = \sum_{T \sim S} \F(T)$. Here $T \sim S$ if and only if $S$ and $T$ differ in one coordinate (that is, are adjacent in $\H$). We have the following important properties of $\F$.

\lem 
\label{lem:F under A}
\begin{itemize}

\item $\sum_{S \in \H} \F(S) = I_m$, where $I_m$ is the $m \times m$ identity matrix.  

\item For any $S \in \H$, we have
\[
(A \F)(S) ~\succeq~ \frac{\l^2}{\l_1} \cdot \F(S).
\]
Here the inequality is in the positive semidefinite ordering. 

\end{itemize}
\elem

\noi This lemma and some other auxiliary claims below will be proved in Section~\ref{sec:proofs}.

\noi We can now proceed with the proof of the theorem. For $x \in \H$, let $W_x$ be the Walsh-Fourier character on $\H$ defined by $W_x(S) = (-1)^{\sum_{i=1}^n x_i S_i}$ for all $S \in \H$. 

\noi Let $C$ be a code of length $n$ and distance $d$. Let the matrix $M$ be defined as follows. The rows of $M$ are indexed by pairs $(y,i)$ where $y$ is an element of $C$ and $1 \le i \le m$. The columns are indexed by the elements of $B$. For $y \in C$, $1 \le i \le m$, and $S \in B$, let $M\big((y,i),S\big) = W_y(S) \cdot \frac{\psi_i(S)}{\phi(S)}$. Let $D$ be the $|B| \times |B|$ diagonal matrix indexed by the elements of $B$, with $D(S,S) = \phi(S)$ for all $S \in B$, and let ${\cal M} = M D M^t$. Then ${\cal M}$ is a $\(|C| \cdot m\) \times \(|C| \cdot m\)$ matrix whose rank is upper bounded by $|B|$. Hence for the proof of the theorem it suffices to show that the rank of ${\cal M}$ is at least $\frac{1}{2d} \cdot |C|m$. For the remainder of the proof we write $N$ for $|C|$, for typographic convenience.

\noi Fix $x$ in $C$, and let $M_x$ be the $m \times |B|$ row submatrix of $M$ with rows indexed by $(x,i)$, $1 \le i \le m$. For $x,y \in C$, let ${\cal M}_{x,y} = M_x D \(M_y\)^t$. Then ${\cal M}_{x,y}$ is an $m \times m$ submatrix of ${\cal M}$ and it is easy to see that ${\cal M}_{x,y} = \sum_{S \in \H}  W_{x+y}(S) \cdot \F(S)$. It seems natural to extend the notion of Fourier transform to matrix-valued functions and to write ${\cal M}_{x,y} = 2^n \cdot \widehat{\F}(x+y)$. Using this notation, we can write ${\cal M} = \big(2^n \cdot \widehat{\F}(x+y)\big)_{x,y \in C}$. 

\noi Let $\l_1...\l_{Nm}$ be the eigenvalues of ${\cal M}$. We will show that $\(\sum_{i=1}^{Nm} \l_i\)^2 \ge \frac{Nm}{2d} \cdot \sum_{i=1}^{Nm} \l_i^2$, or equivalently that 
\beqn
\label{ineq:eigen}
\mathrm{tr}^2\({\cal M}\) ~\ge~ \frac{Nm}{2d} \cdot \mathrm{tr}\({\cal M}^2\).
\eeqn
This will imply (see e.g., \cite{LLC}) that the rank of ${\cal M}$ is at least $Nm / 2d$, proving the claim. In the argument below we will need some simple properties of the extended notion of Fourier transform, which we list in the following lemma. 

\lem 
\label{lem:FT}
Let $M_m(\R)$ denote the space of real $m \times m$ matrices. For functions $F, G:\H \rarrow M_m(\R)$, let $F \ast G:\H \rarrow M_m(\R)$ be defined by $\big(F \ast G\big)(T) = \frac{1}{2^n} \sum_{S \in \H} F(S) G(S+T)$. For $x \in \H$, let $\widehat{F}(x) = \frac{1}{2^n} \sum_{S \in \H} F(S) W_x(S)$.

\noi Then the following holds.
\begin{itemize}


\item $\widehat{F \ast G} = \widehat{F} \cdot \widehat{G}$.

\item Let $A$ be the adjacency matrix of the cube $\H$. Then $\widehat{AF}(x) = (n - 2|x|) \cdot \widehat{F}(x)$.   

\item Let $\mathrm{tr}(B)$ denote the trace of a matrix $B$. Then $\widehat{\mathrm{tr}(F)} = \mathrm{tr}\(\widehat{F}\)$.

\item Let $F, G:\H \rarrow {\S}^m_+$. Then the function $F \ast G$ is not necessarily p.s.d. valued, but we still have (and this is a key point)
\[
\mathrm{tr}\(F \ast G\) ~\ge~ 0.
\] 

\end{itemize}
\elem

\noi We proceed with the proof of (\ref{ineq:eigen}). Let $f(x) =  2^n \cdot \widehat{\F}(x)$. Then $\mathrm{tr}\({\cal M}\) = N \cdot \mathrm{tr}\(f(0)\) = Nm$, where the last step follows from the first claim of Lemma~\ref{lem:F under A}. Note also that $\mathrm{tr}\({\cal M}^2\) = \sum_{x,y \in C} \mathrm{tr}\(f^2(x+y)\)$. Hence, to prove (\ref{ineq:eigen}) we need to show that $\sum_{x,y \in C} \mathrm{tr}\(f^2(x+y)\) \le 2d \cdot Nm$. 

\noi Let $K(x) = \mathrm{tr}\(f^2(x)\)$. Note that $K \ge 0$, since $f(x)$ is a symmetric real matrix. Let $\beta = \widehat{K}$. The following two properties of $\beta$ are simple corollaries of Lemmas~\ref{lem:F under A}~and~\ref{lem:FT}. We have $\beta = 2^{n} \cdot \mathrm{tr}\(\F \ast \F\) \ge 0$ and $A \beta =  2^{n} \cdot \mathrm{tr}\(\F \ast (A\F)\) \ge \frac{\l^2}{\l_1} \cdot \beta$.

\noi In the next calculations we use the notation and some simple notions from the (usual) Fourier analysis on the Hamming cube. We write $\<f,g\> = \frac{1}{2^n} \cdot \sum_{S \in \H} f(S) g(S)$ for the normalized inner product on the cube, and $\<\widehat{f},\widehat{g}\>_{\cal F} = \sum_{x \in \H} \widehat{f}(x) \widehat{g}(x)$ for the (not normalized) inner product of functions in the "Fourier domain", and recall that by the {\it Parseval identity} $\<f,g\> = \<\widehat{f},\widehat{g}\>_{\cal F}$.

\noi Let $C$ be a code with minimum distance $d$. We estimate $\<A\beta,\widehat{1_C}^2\>_{\cal F}$ in two ways. On one hand,
\[
\<A \beta,\widehat{1_C}^2\>_{\cal F} ~\ge~ \frac{\l^2}{\l_1} \cdot \<\beta, ~\widehat{1_C}^2\>_{\cal F} ~\ge~ (n-2d+1) \cdot \<\beta, ~\widehat{1_C}^2\>_{\cal F} ~=~ (n-2d+1) \cdot \<K, ~1_C \ast 1_C\> ~=
\]
\[
\frac{n-2d+1}{2^{2n}} \cdot \sum_{x, y \in C} K(x+y) ~=~ \frac{n-2d+1}{2^{2n}} \cdot \sum_{x, y \in C} \mathrm{tr}\(f^2(x+y)\).
\]
We used the properties of $\beta$ in the first step, the assumption of the theorem in the second step, and Parseval's identity in the third step. On the other hand,
\[
\<A \beta,\widehat{1_C}^2\>_{\cal F} ~=~ \<(n-2|x|) \cdot K(x),~1_C \ast 1_C \> ~=~ \frac{1}{2^{2n}} \sum_{x,y \in C} \Big(n - 2|x+y|\Big) ~K(x+y) ~\le
\]
\[
\frac{n}{2^{2n}} \cdot N K(0) + \frac{n - 2d}{2^{2n}} \cdot \sum_{x \not = y \in C} K\(x + y\) ~=~
\frac{2d}{2^{2n}} \cdot N K(0) ~+~ \frac{n - 2d}{2^{2n}} \cdot \sum_{x, y \in C} K\(x + y\) ~=~ 
\]
\[
\frac{2d}{2^{2n}} \cdot N m ~+~ \frac{n - 2d}{2^{2n}} \cdot \sum_{x, y \in C} \mathrm{tr}\(f^2(x+y)\).
\]

\noi In the first step we used Parseval's identity, in the second step the fact that $\widehat{Ag}(x) = (n-2|x|) \cdot \widehat{g}(x)$ for a function $g$ on the Hamming cube, in the third step the nonnegativity of $K$ and the fact that $C$ has distance $d$, and in the fourth step that $K(0) = m$. Combining the two estimates and simplifying gives (\ref{ineq:eigen}).

\eprf

\noi Let us make a comment about the proof. For $\l = \l_1$, the argument reduces to the argument in the proof of Theorem~1.1 in \cite{LLC}. The key new insight here is to consider functions from the cube to positive semidefinite matrices, which allows to preserve information about eigenspaces of dimension greater than $1$. It turns out that the positive semidefiniteness retains enough positivity for the argument to go through: crucially, although products of positive semidefinite matrices need not be positive semidefinite, their traces are nonnegative. 

\section{Proofs of the remaining claims}
\label{sec:proofs}

\subsubsection*{Proof of Lemma~\ref{lem:F under A}}

\noi The first claim of the lemma follows from the fact that $\psi_1,...,\psi_m$ form an orthonormal basis of $V$ under $\<\cdot,\cdot\>_{\phi}$. Using this fact, it is easy to verify that for any $a \in \R^m$ we have $\<a, \(\sum_{S \in \H} \F(S)\) a\> = \<\sum_{i=1}^m a_i \psi_i, \sum_{i=1}^m a_i \psi_i\>_{\phi} = \sum_{i=1}^m a^2_i$.

\noi We pass to the second claim of the lemma. For $S \not \in B$ the inequality is trivial, since the LHS is a positive semidefinite matrix, while the RHS is zero. Let now $S \in B$. We have
\[
(A \F)(S) ~=~ \sum_{T \sim S} \F(T) ~=~ \sum_{\substack{T \sim S \\ T \in B}} \frac{v(T) \otimes v(T)}{\phi(T)} ~\succeq~ \frac{\(\sum\limits_{\substack{T \sim S \\ T \in B}} v(T)\) \otimes \(\sum\limits_{\substack{T \sim S \\ T \in B}} v(T)\)}{\sum\limits_{\substack{T \sim S \\ T \in B}} \phi(T)} ~=~ \frac{\l^2}{\l_1} \cdot \F(S).
\]
For the last step, note that by the definition of the function $v$, for any $S \in B$ holds $\sum_{\substack{T \sim S \\ T \in B}} v(T) = \l v(S)$. Similarly, 
$\sum_{\substack{T \sim S \\ T \in B}} \phi(T) = \l_1 \phi(S)$. 

\noi It remains to explain the inequality step. Let $L$ be the matrix on the LHS of this inequality, and $R$ the matrix on its RHS. We need to show that the scalar inequality $\<a, La\> \ge \<a, Ra\>$ holds for any $a \in \R^m$ . Writing $s(T)$ for $\<a,v(T)\>$ and $t^2(T)$ for $\phi(T)$ (recalling that $\phi(T) > 0$), the scalar inequality becomes
$\sum_{T} \frac{s^2(T)}{t^2(T)} \ge \frac{\(\sum_{T} s(T)\)^2}{\sum_{T} t^2(T)}$, which is true by the Cauchy-Schwarz inequality.
\eprf

\subsubsection*{Proof of Lemma~\ref{lem:FT}}

\noi The first three claims of the lemma follow directly from definitions, or are easy to verify exactly as in the scalar case. We omit the proofs. 

\noi Here is the argument for the last claim of the lemma. Let $S \in \H$. Then
\[
\mathrm{tr}\Big(\(F \ast G\)(S)\Big) ~=~ \frac{1}{2^n} \sum_{T \in \H} \mathrm{tr}\Big(F(T) G(S+T)\Big) ~\ge~ 0,
\]
since the trace of a product of two positive semidefinite matrices is nonnegative. 
\eprf

\end{document}